\RequirePackage{fix-cm}
\documentclass[twocolumn,epjc3]{svjour3}  
\smartqed  
\RequirePackage{graphicx}
\usepackage{multirow}
\journalname{Eur. Phys. J. C}
\begin{document}

\title{Interaction and Identification of the Di-Hadronic Molecules
}


\author{D. P. Rathaud 
        \and
        Ajay Kumar Rai 
}

\thankstext{e1}{e-mail: dharmeshphy@gmail.com}


\institute{Department of Applied Physics, Sardar Vallabhbhai National Institute of Technology, Surat, Gujarat, India -395007 \label{addr1} 
}

\date{Received: date / Accepted: date}

\maketitle

\begin{abstract}
We study the interesting problem of interaction and identification of the hadronic molecules which seem to be deuteron-like structure. In particular, we propose a binding mechanism in which One Boson Exchange Potential plus  Yukawa screen-like potential is applied in their relative s-wave state. We propose the dipole-like interaction between two color neutral states to form a hadronic molecule.  For the identification of the hadronic molecules, the Weinberg's compositeness theorem is used to distinguish the molecule from confined (elementary) state.  The present formalism predict some di-hadronic molecular states, involving quarks (s, c, b or $\overline{s}$, $\overline{c}$, $\overline{b}$) as a constituents, namely, $pn$,  $K\overline{K}$, $\rho \overline{\rho}$, $K^{*}\overline{K^{*}}$, $D\overline{D^{*}}$($\overline{D}D^{*}$), $D^{*}\overline{D^{*}}$, $B\overline{B^{*}}$, $B^{*}\overline{B^{*}}$,
 $D^{*\pm}\overline{D_{1}^{0}}$, $ D^{0}\overline{K^{\pm}}$, $D^{*0}\overline{K^{\pm}}$, with their possible quantum numbers.
\end{abstract}

\section{Introduction}
\label{}

From few decades, tremendous efforts have been made for search of the hadronic molecules on both the theoretical and experimental forefront. However, apart from deuteron, we are still waiting for another strong molecular candidates, which are expected in the fundamental theory(QCD).  In the last few years, there have been several experimental discoveries of manifestly narrow exotic resonances X(3872) \cite{Choi-prl2003-91}, $Z(4430)^{+}$ \cite{Choi-prl2008-100},Y(4260) \cite{Coan-prl2006-96}, $Z_{b}(10610)/(10650)$ \cite{Chen-prl2008-100}, $P_{c}(4450)$ \cite{Aaij-prl2015-115-pentaquark} and many more.

The various theoretical model have been proposed to explain these exotic states, such as, hadronic molecule \cite{Guo-RevModPhys2018-90,Geng-prd2009-79,Martinez-plb2013-719,Molina-prd2008-78,Garca-Recio-prd2013-87}, conventional quarkqonia \cite{WANG-plb2011,Fulvia-prd2009,Ferretti-prd2014,Kalashnikova-prd2010}, compact tetra-quark and pentaquark \cite{Ali-Prog.Part.Phys2017-97,Esposito-phys.Rep2017-668}, cups like effects\cite{Guo-prd2015-92}. Indeed, these efforts have been made with various theoretical approaches like effective field theory \cite{Hidalgo-Duque-prd2013,Fleming-prd2007,Guo-epjc2014,Miguel-epjc2015}, QCD sum rule \cite{J.R.Zhang-prd2012-86,Azizi-prd2017-95,Azizi-prd2017-96}, lattice QCD \cite{Alford-npb2000-578,Meng-prd2009-80,L.Liu-prd2013-87,Bali-prd2017-96}, gauge invariant model \cite{Branz-prd2008-78}, potential model \cite{Weinstein-prd1983-27,Weinstein-prd1990-41,Weinstein-prd1993-47,Thomas-prd2008,Hanhart-prd2007-75,Zhang-prd2002-65,Ding-prd2009,Iwasaki-prd2008-77,Rai-epjc2015}. One can find brief review on subject in Refs. \cite{Guo-RevModPhys2018-90,Esposito-phys.Rep2017-668,Ali-PPNP2017-2017,Swanson-PhysRep2006-429,Klempt-PhysRept2007-454,Olsen-RevModPhys2018-90}

The present article focused on the hadronic molecular model. We have studied  hadronic molecules within the potential model framework. There are two prerequisites for the study molecular model (i) interaction between two color neutral hadron, through which two hadrons can formed hadronic molecule (ii) the identification of the molecular state from others. 

In 1991, T$\ddot{o}$rnqvist \cite{Tornqvist-prl1991} proposed the deuteron-like loosely bound mesonic molecules meditated by One Pion Exchange (OPE). Ericson and Karl \cite{Ericson-Plb1993} questioned on the strength of the OPE potential and put forward a reasonable argument that the critically minimal reduced mass is required for the formation of the molecule. Indeed, it was also noted that the OPE potential alone could not gain sufficient attractive depth for the molecular binding \cite{Tornqvist-prl1991,Tornqvist-zphysc1994,Ericson-Plb1993,Swanson-PhysRep2006-429,Esposito-phys.Rep2017-668}. The OPE potential describe the long range interaction while short range interaction still being unexplained. In Ref. \cite{Swanson-PhysRep2006-429,Barnes-prd1992-46}, authors have assumed  the interaction between mesonic molecule as an instantaneous confining interaction and a short range spin-dependent interaction motivated by one gluon exchange.  But, from this quark level interaction, one could not distinguished the difference between molecular and compact tetraquark states. 
Certainly, in the phenomenological study of molecular model, one required a reliable molecular interaction potential to get shallow bound state which also give the clue for the dynamics of decays.  In Ref. \cite{Esposito-phys.Rep2017-668} authors have discussed the open problem with molecular model and suggested di-quark-di-anitquark interaction for explanation of these XYZ state. 
 
Therefor, the subject of the molecular model is required more efforts from both theoretical and experimental point of view. Following the motivation of molecular model and interaction potential, as well as, to  scanning the internal structure of these resonances and looking towards the molecular interpretation, we need to know both the two prerequisite discussed above.  Henece, in the present paper, both these two prerequisites are attempted.   
\subsection*{Inter-hadronic interaction}
Deuteron as a bound state of the proton and neutron has been extensively studied state, and it has well established status \cite{Machleidt-PhysRep1987-149,Machleidt-prc2001-63,Naghdi-PhysParNucl2014-45,Paris-prc2000-62}. Therefor, it is preferable to take deuteron as a model for the study and predictions of other deuteron-like hadronic molecules. For such bound state, there are two question arises- (a) how two hadrons are forming bound state? and (b) why two color neutral hadrons attract each others? According to  definition of the molecule, a sufficient attractive potential strength has needed to get a bound state and this strength could parameterized by an effective coupling constant. Indeed, in the relation to the first question, `how'- various realistic potentials like full-Bonn (CD-Bonn), Nijmegen-Group of, Paris-Group of potential etc.. have been developed (see Ref.\cite{Naghdi-PhysParNucl2014-45} for review). The exchange of particles has basis for all these realistic potentials and to be known as a One Boson Exchange (OBE) potential; exploring the initial idea of Yukawa's pion exchange interaction to explain the nuclear force. In the OBE, long, mid and short distance interaction have been introduced through exchange of mesons where the range of the force depend on the mass of the exchange mesons. To give a precise answer for the second question that why two color neutral hadron makes bound system being difficult. Because, it requires very elegant knowledge of fundamental interaction at very short distance where the OBE has limitation to explain such interaction. 

In this paper, we propose the dipole-dipole like interaction between two color neutral hadrons, which could be either permanent dipole or induced dipole in which the latter one is weakest. The uneven color charge field distribution between two hadrons creates a dipole effect while in other case the color charge field of two hadron get influence to each other and leads to temporary dipole. On such proposal we would note two points  (i) the strongly increasing strength of the interaction leads two hadronic states to bare confining state (ii) the attraction and repulsion is depending on respective aliment of the dipoles in spin-isospin space.  

For the calculation of the s-wave mass spectra of deuteron-like di-hadronic systems, we have used the s-wave OBE potential plus Yukawa screen-like potential, where Yukawa screen-like potential represent the dipole-like interaction.  Thus, the condition for the existence of the molecular system are: (i) the kinetic energy of the system must be less i.e. the hadrons should be heavy enough to get bound state (ii) the two hadron should carry lowest orbital angular momentum ($l$=0) (iii) molecular state should be loosely bound (i.e. below threshold) and narrow.  Here, we emphasize that the s-wave OBE plus screen type potential could apply, in principle, for binding mechanism between two hadrons. 
\subsection*{Compositeness theorem}
As we mention that the second prerequisite or challenge for the hadronic molecular model is the identification of the molecular state from the bare elementary state. Thus, to attempt this prerequisite, we used the Weinberg's compositeness theorem \cite{Weinberg-pr1965-137}. 

In the Sixties, Weinberg \cite{Weinberg-pr1965-137} suggested in a sophisticated way that the deuteron were a composite particle. In his novel work, he tried to show an elegant  model-independent  way to identify whether a particle is in a bare elementary state or in a composite state. The conclusion was based on a generalization of Levinson's theorem which gives the formulas for scattering length $a_{s}$ and effective range $r_{e}$ in terms of Z, where Z is the ``field renormalization" constant \cite{Weinberg-pr1965-137}, 
\begin{eqnarray}
\label{as and re equation}
&&a_{s}=\left[2(1-Z)/(2-Z)\right]R+{\cal{O}}(1/\beta) \nonumber \\
&&r_{e}=\left[-Z/(1-Z)\right]R+{\cal{O}}(1/\beta)
\end{eqnarray}
where  $R\equiv 1/ \sqrt{2\mu \epsilon}$, $\epsilon$ is the binding energy and $\mu$ is the reduced mass of the composite system. The ${\cal{O}}(1/\beta)$ is the range of the force and could be calculated if one know the information of the interaction. In order to determine the state of the particle as in a bare elementary or in a composite state, he argued that the renormalization constant Z takes the value  $0 \leq Z \leq 1$. If Z=0 then the particle is in a pure composite state, while for Z=1 it becomes a purely elementary. This argument have previously discussed by other authors \cite{Vaughn-PhesRev1961,Acharya1962} and followed by Weinberg\cite{Weinberg-pr1965-137}. For the case Z=0 (the deuteron as a composite particle) the Eq(\ref{as and re equation}) becomes $a_{s}=R$ and $r_{e}={\cal{O}}(1/\beta)$ which is in agreement with the experimental vales : $a_{s}=+5.41$ fm, $r_{e}=+1.75$ fm. 

In addition, he were made remarked for such investigation as\cite{Weinberg-pr1965-137}; (i) the composite particle must couple to a two-particle channel with threshold not too much above the composite particle mass (ii) composite particle must be in a s-wave (l=0)(iii) composite particle must be stable. In the case of deuteron all these three conditions are all most satisfied. Deuteron followed the third condition very well due to the lightest baryons in its internal substructure which preserves the stability of the deuteron. Whereas, it is not possible in all other deuteron-like cases. Usually in other cases, one could get (i) and (ii) satisfied. V. Baru et. al. \cite{Baru-plb2004-586} have applied this formalism to study the state $a_{0}(980)$ and $f_{0}(980)$ in the light meson sector. Recently, Xian-Wei Kang and J. A. Oller \cite{Kang-epjc2017-77} have extended and advance this theorem to analyze X(3872) and introduce a near-threshold parameterization. \\
 
The aim of the present study is to take an attention on some interesting possibilities for molecular (deuteron-like) structure with proposed interaction and identification with observables. Hence, we approximate the interaction potential as discussed and used the  Weinberg's \cite{Weinberg-pr1965-137} approach in order to determine a composite state from a bare state. 

In the present paper, we have fitted the potential parameters to get the deuteron binding energy and used same for rest of the calculation. In such a way, we have fixed the model for di-hadronic calculation. We have noticed that the s-wave OBE potential could not get sufficient attractive strength to get bound state, not even in attractive spin-isospin channels, therefor, the screen Yukawa-like potential is being used for additional attractive strength. 
The mass spectra of di-hadronic (di-mesonic, meson-baryon and di-baryonic) states are calculated by using proposed interaction. In this article, the mass spectra of the di-mesonic states are presented. The mass spectra of meson-baryon and di-baryon molecules along with detail analysis of meson exchange interaction potential will presented in the separate publication.

The article is organized as follows: after the brief introduction, the effective interhadronic potential  and theoretical framework are discussed in section-2. The results are presented and discussed in the section-3 and finally summary and conclusion of the work are presented in the last section of the article.

\section{Effective Potential (OBE+Yukwa-like Screen)}
\label{}
Let us describe the interaction potential in terms of the s-wave One Boson Exchange (OBE) potential and   phenomenological attractive screen Yukawa-like  potential. 

The light mesons under consideration for the OBE Potential are as follows \cite{Machleidt-PhysRep1987-149,Machleidt-prc2001-63}: Pseudoscalar meson $(ps)=\pi,\eta$ ; Scalar meson $(s)=\sigma,\delta$ and Vector meson $(v)=\omega,\rho$. The OBE potential is the sum of the all one meson exchange, namely
\begin{eqnarray}
V_{OBE}=V_{ps}+V_{s}+V_{v}
\end{eqnarray}
where the individual s-wave one meson exchange interaction potential expressed as  \cite{Machleidt-PhysRep1987-149} 
\begin{eqnarray}
&V_{ps}&=\frac{1}{12}\left[\frac{g^{2}_{\pi qq}}{4\pi}\left(\frac{m_{\pi}}{m}\right)^{2} \frac{e^{-m_{\pi}r_{ij}}}{r_{ij}} \right. \nonumber \\
&& \left. \left(\tau_{i}\cdot\tau_{j}\right)+\frac{g^{2}_{\eta qq}}{4\pi}\left(\frac{m_{\eta}}{m}\right)^{2} \frac{e^{-m_{\eta}r_{ij}}}{r_{ij}}\right]\left(\sigma_{i}\cdot\sigma_{j}\right)
\end{eqnarray}
\begin{eqnarray}
&V_{s}&=-\frac{g_{\sigma qq}^{2}}{4\pi}m_{\sigma}\left[1-\frac{1}{4}\left(\frac{m_{\sigma}}{m}\right)^{2}\right]\frac{e^{-m_{\sigma}r_{ij}}}{m_{\sigma}r_{ij}}+ \nonumber \\ &&\frac{g_{\delta qq}^{2}}{4\pi}m_{\delta}\left[1-\frac{1}{4}\left(\frac{m_{\delta}}{m}\right)^{2}\right]\frac{e^{-m_{\delta}r_{ij}}}{m_{\delta}r_{ij}}\left(\tau_{i}\cdot\tau_{j}\right)
\end{eqnarray}
\begin{eqnarray}
&V_{v}&=\frac{g_{\omega qq}^{2}}{4\pi}\left(\frac{e^{-m_{\omega}r_{ij}}}{r_{ij}}\right)+ \nonumber \\
&& \frac{1}{6}\frac{g_{\rho qq}^{2}}{4\pi}\frac{1}{m^{2}}\left(\tau_{i}\cdot\tau_{j}\right)\left(\sigma_{i}\cdot\sigma_{j}\right)\left(\frac{e^{-m_{\rho}r_{ij}}}{r_{ij}}\right)
\end{eqnarray}

The OBE potential with finite size effect due to extended structure of the hadrons can be expressed as \cite{Machleidt-PhysRep1987-149}
\begin{eqnarray}
V_{\alpha}(r_{db})=V_{\alpha}(m_{\alpha},r_{db})- &&F_{\alpha 2}V_{\alpha}(\Lambda_{\alpha 1},r_{db})\nonumber \\ && +F_{\alpha 1}V_{\alpha}(\Lambda_{\alpha 2},r_{db})
\end{eqnarray}
where $\alpha$ = $\pi$, $\eta$, $\sigma$, $\delta$, $\omega$ and $\rho$ mesons, while  
\begin{eqnarray}
\Lambda_{\alpha 1}=\Lambda_{\alpha}+\epsilon \hspace{0.99cm} and \hspace{0.99cm} \Lambda_{\alpha 2}=\Lambda_{\alpha}-\epsilon \nonumber \\ 
F_{\alpha 1}=\frac{\Lambda_{\alpha 1}^{2}-m_{\alpha}^{2}}{\Lambda_{\alpha 2}^{2}-\Lambda_{\alpha 1}^{2}} \hspace{0.5cm} and \hspace{0.5cm} F_{\alpha 2}=\frac{\Lambda_{\alpha 2}^{2}-m_{\alpha}^{2}}{\Lambda_{\alpha 2}^{2}-\Lambda_{\alpha 1}^{2}}\\ \nonumber
\end{eqnarray}
the subscript $\alpha$ tends for mesons ($\pi$, $\eta$, $\sigma$, $\delta$, $\omega$ and $\rho$)  $\epsilon/\Lambda_{\alpha}\ll1$, thus $\epsilon$=10 MeV is an appropriate choice \cite{Machleidt-PhysRep1987-149}.

The overall contribution form s-wave OBE is very less due to its delicate cancellation of the individual one meson exchange contribution with each other. Here, we want to make two remarks  on the overall contribution (attraction/repulsion) of the s-wave OBE potential: (i) its contribution is strongly related to the coupling constant of the each individual meson exchange and (ii) it is depends on the spin-isospin channels. 

\begin{table}[]
\begin{center}
\caption{OBE potential parameters, this parameters are taken from \cite{Machleidt-PhysRep1987-149,Machleidt-prc2001-63}}
\label{OBEP parameters}
\scalebox{0.85}{
\begin{tabular}{ccccccc}
\hline
\hline
Mesons & $\pi$ & $\eta$ & $\sigma$  & $a_{0}(\delta)$ & $\omega$ & $\rho$ \\
\\
$\frac{g^{2}_{\alpha NN}}{4\pi}$ & 13.6 & 3 & 7.7823 $^{*}$ & 2.6713 & 20 & 0.85 \\
\\
$\Lambda_{\alpha}$ & 1.3 & 1.5 & 2.0 & 2.0 & 1.5 & 1.3\\
\\
Mass (in MeV) & 134.9 & 548.8 & 710 & 983 & 782.6 & 775.4 \\
\hline
\hline
\end{tabular}
}
\end{center}

\begin{center}
{\tiny($^{*}$The  $\frac{g^{2}_{\sigma NN}}{4\pi}$ for the $\sigma$-exchange given in the table is used for total isospin $I_{T}$=1. Whereas for $I_{T}$=0, $\frac{g^{2}_{\sigma NN}}{4\pi}$=16.2061 have been used.)}
\end{center}
\end{table} 
The masses of exchange mesons, coupling constant and the regularization parameter($\Lambda_{\alpha}$) are tabulated in Table-\ref{OBEP parameters}.
The estimates of the coupling constant are given in the most of the realistic potentials \cite{Machleidt-PhysRep1987-149,Machleidt-prc2001-63,Naghdi-PhysParNucl2014-45} which are developed to reproduce NN-phase shift data and explain the deuteron properties. We have taken them same as estimated in Refs. \cite{Machleidt-PhysRep1987-149,Machleidt-prc2001-63} and approximated the meson-hadron coupling constant for other hadronic molecular cases as 
\begin{equation}
g_{\alpha hh} \simeq g_{\alpha NN}
\end{equation}
where $g_{\alpha hh}$ and $g_{\alpha NN}$ are the meson-hadron and meson-nucleon coupling constants, respectively.

The strength of effective s-wave OBE is vary shallow due to very delicate cancellation of the individual meson exchange with each other, therefore, we have incorporated the Yukawa-like screen potential to get additional strength for net effective di-hadronic interaction potential. Screen Yukawa-like potential is incorporated, namely
  
\begin{eqnarray}
V_{Y}&=& -\frac{k_{mol}}{r_{ij}} e^{\frac{{-c^{2}r_{ij}^{2}}}{2}}
\end{eqnarray}

here, $k_{mol}$ is the residual running coupling constant and c is a screen fitting parameter.  $k_{mol}$ can be estimated by using formula,
\begin{equation}
\label{Kmol}
k_{mol}(M^{2}) = \frac{4\pi}{(11-\frac{2}{3}n_{f})ln\frac{M^{2}+ {M_{B}}^{2}}{\Lambda_{Q}^{2}}}
\end{equation}
where M=2$m_{d}$ $m_{b}$/ ($m_{d}$+$m_{b}$), $m_{d}$ and $m_{b}$ are constituent masses, $M_{B}$=1 GeV, $\Lambda_{Q}$ is taken 0.413 GeV and 0.250 GeV for light and heavy mesons, respectively. The term $n_{f}$ is number of flavour \cite{Ebert-prd2009,Badalian-prd2004}.

The net inter hadronic interaction potential $V_{hh}$ is given as  
\begin{eqnarray}
\label{effective net potential}
V_{hh}=V_{OBE}+V_{Y}
\end{eqnarray}
The masses, exchange meson coupling constant and $\Lambda_{\alpha}$ are the fixed parameters  and obtained from Refs. \cite{Patrignani-PDG2016,Machleidt-PhysRep1987-149,Machleidt-prc2001-63}, also tabulated in Table-\ref{OBEP parameters}. The residual running coupling constant $k_{mol}$ is calculated by using Eq.(\ref{Kmol}). The color screening parameter 'c' is the only free parameter of the model and we fitted it to get the experimental value of binding energy of the deuteron. 

For c=0.0686 GeV, we obtained the binding energy of the deuteron. Hence, we took it as a constant and have not changed for any further calculations of the di-hadronic molecules.

The Hamiltonian of di-hadronic molecule express as 
\begin{equation}
H=\sqrt{P^2+m_{h1}^{2}}+\sqrt{P^2+m_{h2}^2}+V_{hh}
\end{equation}
here, $m_{h1}$ and $m_{h2}$ are the masses of constituent and P is the relative
momentum of two hadrons while the $V_{hh}$ is the inter hadronic interaction potential. 

Within variational approach, we use the hydrogenic trial wave function to determine the expectation value of the Hamiltonian, namely
\begin{eqnarray}
H\psi &=& E\psi \nonumber \\  & \nonumber and & \\
\left\langle K.E.\right\rangle &=& \frac{1}{2} \left\langle \frac{r_{ij}dV_{hh}}{dr_{ij}} \right\rangle
\end{eqnarray}
The variational parameter ($\mu$) is determined for each state by using the Virial theorem.

one should note that the pseudoscalar exchanges are not allowed between two pseudoscalar systems due to parity conservation, as three pseudoscalar at one vertex do not conserved parity, while the parity is good quantum number and well conserved in the strong interaction. Therefore, in the case of $K\overline{K}$ and $D^{0}\overline{K^{\pm}}$ dimesonic states, only $\sigma$, $a_{0}$(or $\delta$) and $\omega$ exchanges contributes to OBE potential. 

In the calculation of di-mesonic states, the $m_{\sigma}$=750 MeV is taken for total isospin $I_{T}$=0 and $m_{\sigma}$=550 MeV is taken for $I_{T}$=1. For  pseudoscalar-pseudoscalar states $f_{0}(980)$, $a_{0}(980)$ and $D_{s0}(2317)^{\pm}$, only the $\sigma$, $a_{0}$ (or $\delta$), and $\omega$ exchange contributes to net s-wave OBE potential, hence, only these meson exchange are considered.

The obtained mass spectra of di-mesonic states  presented in the next section.
\begin{center}
\begin{figure*}
\caption{The characteristic behavior of effective s-wave OBE potential with range is shown in this figure. The graphs are plotted for respective spin-isospin channels of attempted di-mesonic states}
\label{s-wave obe plots}
\includegraphics[scale=0.45]{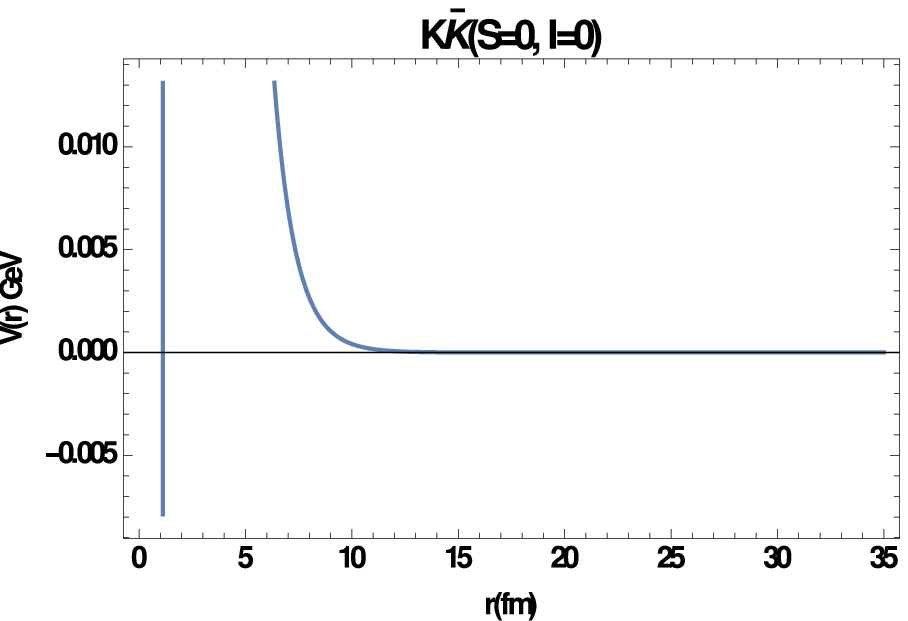}
\includegraphics[scale=0.45]{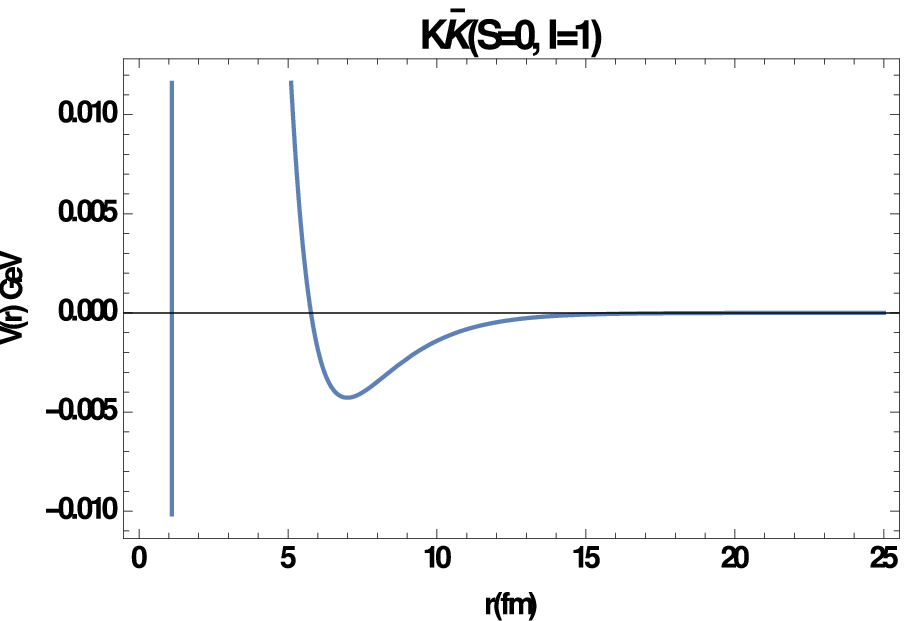}
\includegraphics[scale=0.45]{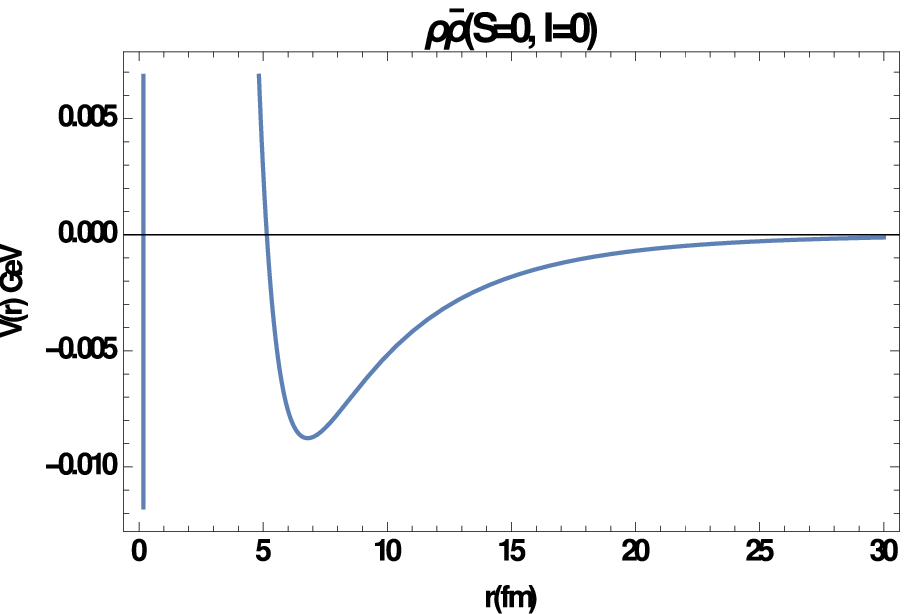}
\includegraphics[scale=0.45]{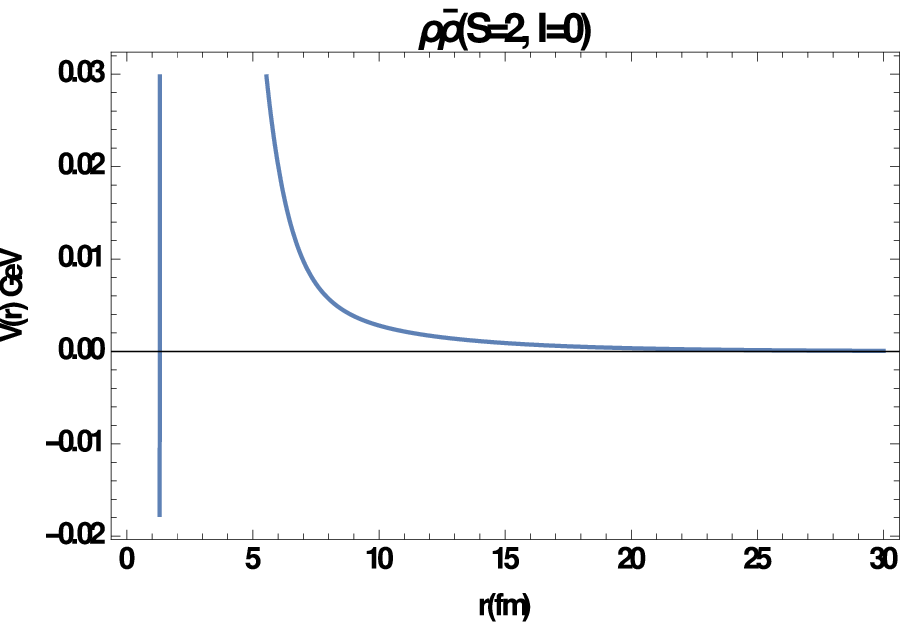}
\includegraphics[scale=0.45]{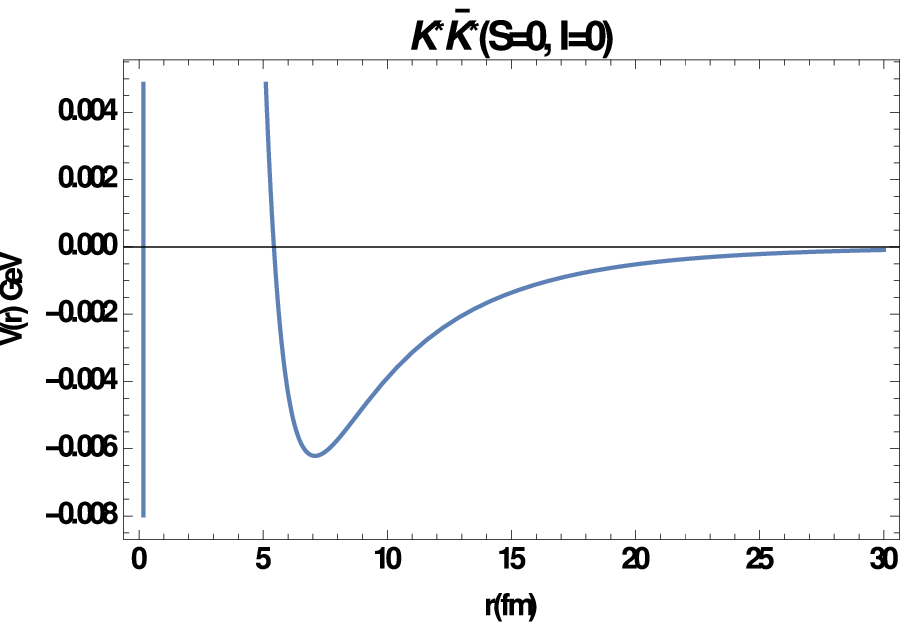}
\includegraphics[scale=0.45]{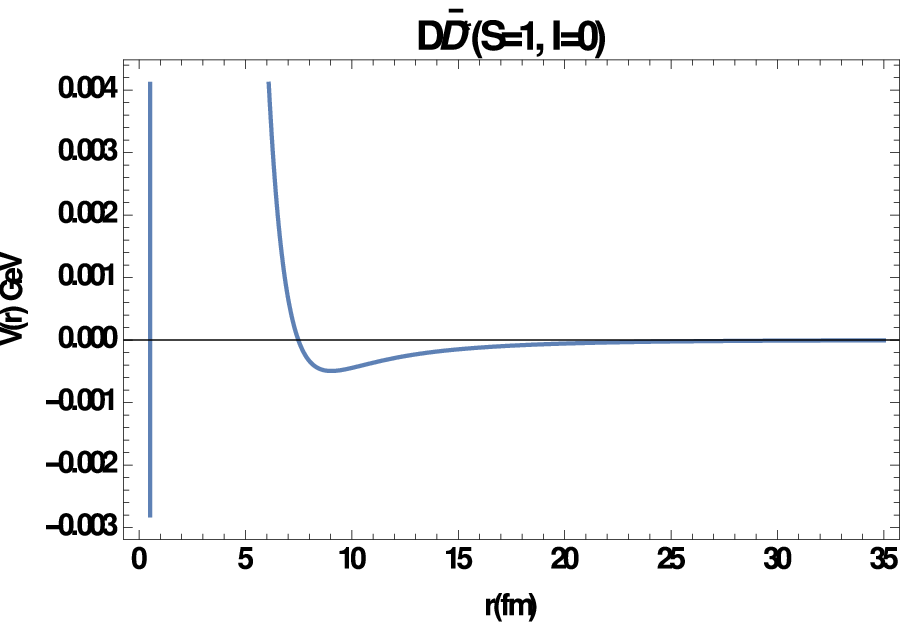}
\includegraphics[scale=0.45]{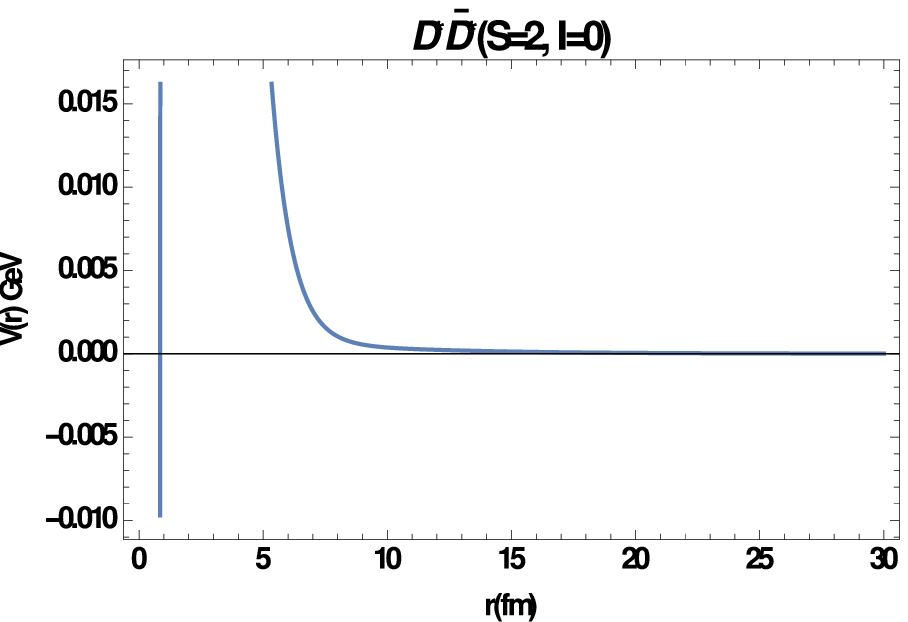}
\includegraphics[scale=0.45]{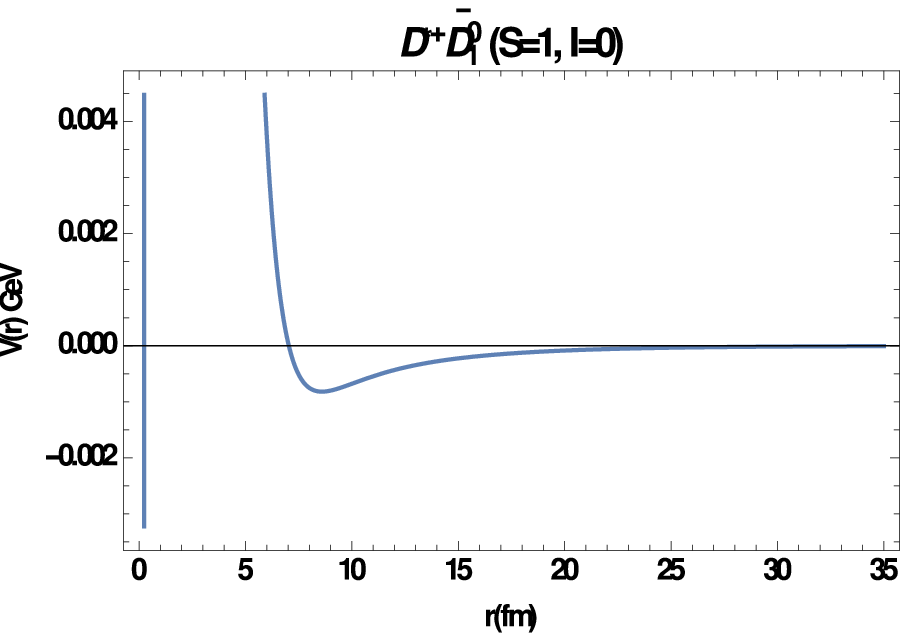}
\includegraphics[scale=0.45]{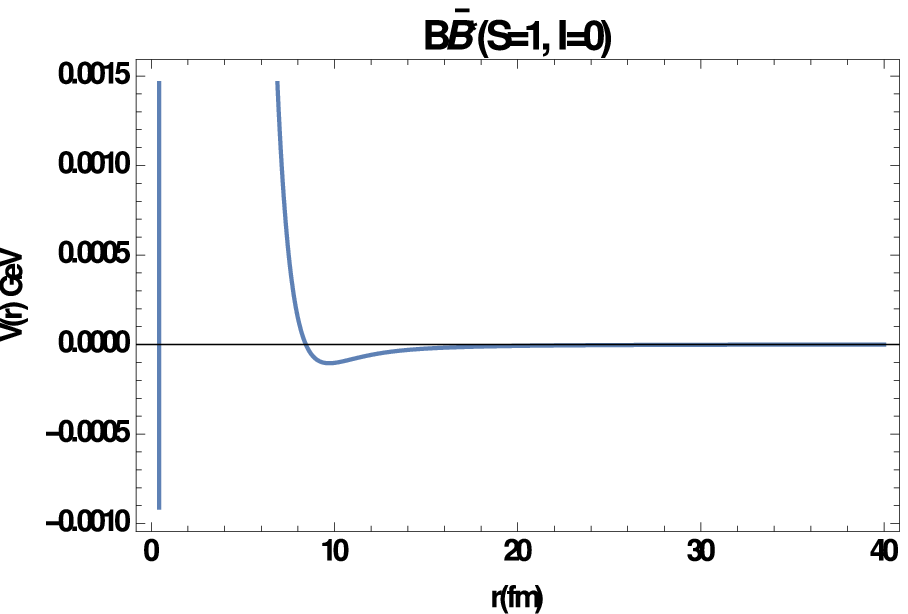}
\includegraphics[scale=0.45]{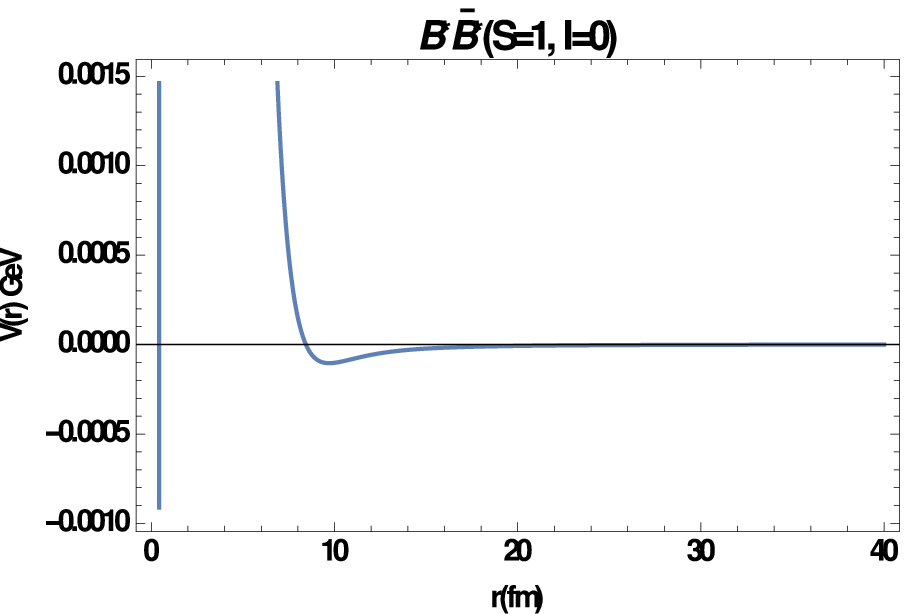}
\includegraphics[scale=0.45]{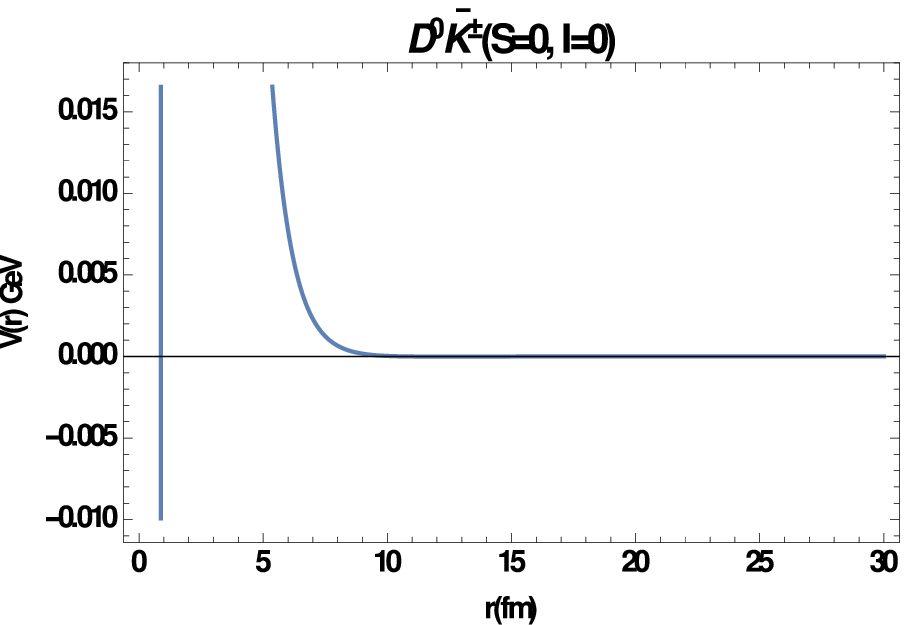}
\includegraphics[scale=0.45]{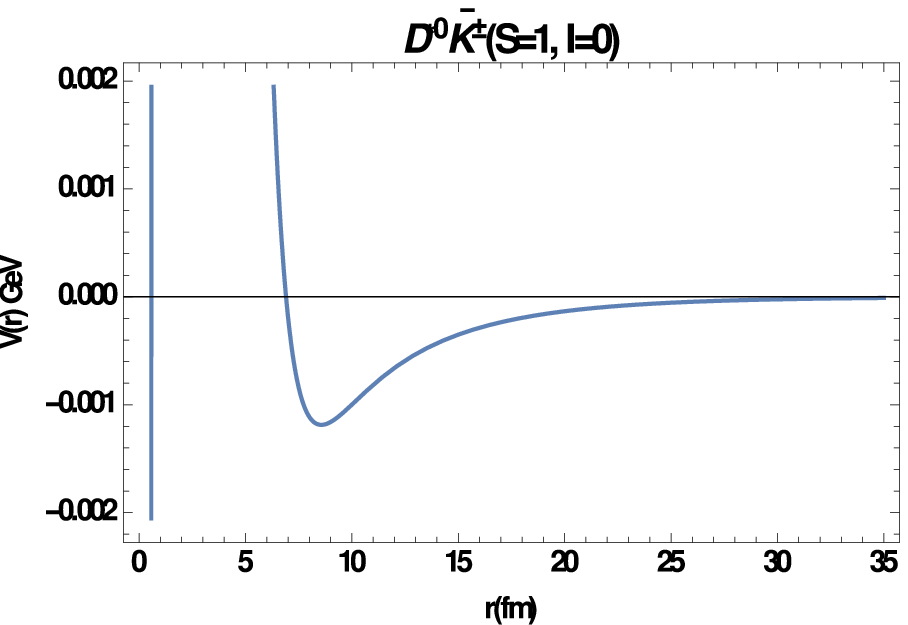}
\end{figure*}
\end{center}


\begin{table*}
\begin{center}
\caption{This table represents some mesonic states which are predicted as di-mesonic molecules in the literatures. According to respective dimesonic state, the threshold mass, reduce mass and natural energy scale of bound states are presented. }
\label{natural energy scale}
\begin{tabular}{ccccc}
\hline
States & Molecular& Threshold & Reduce & Natural energy   \\
& interpretation & mass & mass ($\mu$) &scale($m_{\pi}^{2}/2\mu$)\\
& & MeV & MeV & MeV\\
\hline
\hline 

Deuteron & $pn$ & 1877.84  & 469.45 &19.40  \\

$f_{0}(980)$& $K\overline{K}$ &  995.228 & 248.80 & 36.61  \\

$a_{0}(980)$& $K\overline{K}$ &  995.228 &  248.80&36.61  \\

$f_{0}(1500)$& $\rho\overline{\rho}$ &  1550.98 & 387.74 &23.49 \\

$f^{'}_{2}(1525)$& $\rho\overline{\rho}$ &  1550.98 &387.74 &23.49 \\

$f_{0}(1710)$& $K^{*}\overline{K^{*}}$ & 1791.88 &447.97 &20.33 \\

$X(3872)$ & $D\overline{D^{*}}$ &  3871.84 &966.65 &09.42  \\

$X_{2}(4013)$ & $D^{*}\overline{D^{*}}$ &  4013.96 & 1003.49 &09.07 \\
$Z(10610)$& $B\overline{B^{*}}$ &  10604.78 &  2651.1&03.43 \\
$Z(10650)$& $B^{*}\overline{B^{*}}$ &  10650.4 & 2662.2 &03.42 \\

$Z(4430)^{+}$ & $D^{*+}\overline{D_{1}^{0}}$ & 4431.58 &1098.3 &08.29 \\

$D_{s0}(2317)^{\pm}$ & $D^{0}\overline{K^{\pm}}$ & 2358.53 & 390.34 &23.33 \\

$D_{s1}(2460)^{\pm}$ & $D^{*0}\overline{K^{\pm}}$ &  2500.65 & 396.21 & 22.57 \\
\hline
\hline
\end{tabular}
\end{center}
\end{table*}

\begin{table*}[]
\caption{\small The threshold of hadronic molecules as well as expected binding energy in comparison of receptive exotic states are presented. The calculated binding energy, mass and root mean square radius with possible S-wave quantum numbers are presented and compared with possible exotic states. All hadron masses are taken from PDG \cite{Patrignani-PDG2016}}
\label{threshold and expected binding energy}
\begin{center}
\begin{tabular}{cccccccccc}
\hline
Candidate & Molecular &s-wave & Threshold & Exp. & Expected &\multicolumn{2}{c}{This Work}& rms \\
& interpretation &I$(J^{P})$ & mass & mass  & B.E. & B.E. & mass  & $\sqrt{r^{2}}$  \\
&&&MeV&MeV&MeV&MeV&MeV&fm\\
\hline
\hline
\\
Deuteron & $pn$ & $0(1^{+})$& 1877.84  & $1875.6$ & -2.224 & -2.221 & 1875.5 & 03.13 \\

$f_{0}(980)$& $K\overline{K}$ & $0(0^{+})$ & 995.228 & $990\pm 20$ & -5.228 & -6.154 & 989.07 & 02.11  \\

$a_{0}(980)$& $K\overline{K}$ & $1(0^{+})$ & 995.228 & $980\pm 20$ & -15.228 & -8.656 & 986.57 & 02.01  \\

$f_{0}(1500)$& $\rho\overline{\rho}$ & $0(0^{+})$ & 1550.98 & $1505 \pm 6$ & -45.98 & -5.483 & 1545.5 & 03.05 \\

$f^{'}_{2}(1525)$& $\rho\overline{\rho}$ & $0(2^{+})$ & 1550.98 & $1525 \pm 5$ & -25.98 & +0.003 & 1550.98 & 61.72 \\

$f_{0}(1710)$& $K^{*}\overline{K^{*}}$ & $0(0^{+})$ & 1791.88 & $1722\pm 5$ & -69.88  & -6.853 &  1785.03 & 02.75  \\

$X(3872)$ & $D\overline{D^{*}}$ & $0(1^{+})$ & 3871.84 & $3871.68\pm 0.17$ & -0.160 & -2.395 & 3869.4 & 03.04 \\

$X_{2}(4013)$ & $D^{*}\overline{D^{*}}$ & $0(2^{+})$ & 4013.96 & - & - & -3.862 & 4010.1 &  02.44 \\
$Z(10610)$& $B\overline{B^{*}}$ & $0(1^{+})$ & 10604.78 & $10607.2\pm 2.0$ & +2.42 & -2.295 & 10602.7 & 02.99 \\
$Z(10650)$& $B^{*}\overline{B^{*}}$ & $0(1^{+})$ & 10650.4 & $10652.2\pm 1.5$ & +1.80 & -2.288 & 10648.1  & 02.99  \\

$Z(4430)^{+}$ & $D^{*+}\overline{D_{1}^{0}}$ & $0(0^{-})$ & 4431.58 & $4430$ & -1.580 & -2.656 & 4425.6 & 03.01 \\

$D_{s0}(2317)^{\pm}$ & $D^{0}\overline{K^{\pm}}$ & $0(0^{+})$ & 2358.53 & $2318\pm10$ & -40.53 & -15.63 & 2342.9 & 01.75  \\

$D_{s1}(2460)^{\pm}$ & $D^{*0}\overline{K^{\pm}}$ & $0(1^{+})$ & 2500.65 & $2459.6\pm0.9$ & -41.05 & -12.26 & 2488.3 & 02.08  \\


\hline
\hline 
\end{tabular}
\end{center}
\end{table*}   

\begin{center}
\begin{table*}[]
\caption{The scattering length $a_{s}$ and effective range $r_{e}$ are calculated for di-hadronic systems by using Eq.(\ref{as and re equation}) for different values of renormalization constant Z. The expected (experimental) binding energy is considered in the calculation, also shown in Table-\ref{threshold and expected binding energy}. The values of $a_{s}$ and $r_{e}$ are in fm.}
\label{expected as and re from exp BE}
\scalebox{0.85}{
\begin{tabular}{cccccccccccccccc}
\hline
State   &\multicolumn{2}{c}{Z=0} &\multicolumn{2}{c}{Z=0.2} &\multicolumn{2}{c}{Z=0.4} &\multicolumn{2}{c}{Z=0.5} &\multicolumn{2}{c}{Z=0.6} &\multicolumn{2}{c}{Z=0.9} &\multicolumn{2}{c}{Z=1} \\
\cline{2-15}
 &$a_{s}$ &$r_{e}$ &$a_{s}$ &$r_{e}$& $a_{s}$ &$r_{e}$ &$a_{s}$ &$r_{e}$ & $a_{s}$ &$r_{e}$ & $a_{s}$ &$r_{e}$ & $a_{s}$ &$r_{e}$\\
\hline
\hline
\\
Deuteron &5.78 & 1.46 & 5.3 & 0.38 & 4.7 & -1.42 & 4.34 & -2.86 & 3.93 & -5.01 & 2.9 & -15.81 & 1.46 &-\\
$f_{0}(980)$& 4.13 & 0.26 & 3.7 & -0.7 & 3.16 & -2.32 & 2.84 & -3.61 & 2.47 & -5.54 & 1.55 & -15.21 & 0.26 &-\\
$a_{0}(980)$& 2.63 & 0.36 & 2.37 & -0.21 & 2.06 & -1.15 & 1.87 & -1.91 & 1.65 & -3.04 & 1.11 & -8.71 & 0.36 & -\\
$f_{0}(1500)$ & 2.51 & 1.46 & 2.39 & 1.2 & 2.25 & 0.77 & 2.16 & 0.42 & 2.06 & -0.11 & 1.81 & -2.72 & 1.46 & - \\
$f_{2}(1525)$ & 2.85 & 1.46 & 2.7 & 1.11 & 2.5 & 0.54 & 2.39 & 0.07 & 2.26 & -0.62 & 1.93 & -4.1 & 1.46 & - \\
$f_{0}(1710)$ & 2.25 & 1.46 & 2.16 & 1.26 & 2.05 & 0.94 & 1.99 & 0.67 & 1.91 & 0.28 & 1.72 & -1.69 & 1.46 & - \\
$X(3872)$ & 12.68 & 1.46 & 11.44 & -1.34 & 9.88 & -6.02 & 8.94 & -9.76 & 7.87 & -15.37 & 5.2 & -43.42 & 1.46 & - \\

\multirow{2}{*}{$Z(10610)$} & 1.46  & 1.46 & 1.46  & 1.46  & 1.46  & 1.46  & 1.46  & 1.46 & 1.46  & 1.46  & 1.46  & 1.46  & 1.46& \multirow{2}{*}{-}\\
&- 1.7i & 0.0i&– 1.5i &+ 0.4i & - 1.3i & + 1.1i & - 1.1i & + 1.7i & - 0.9i & + 2.6i & - 0.5i & + 6.9i &0.0i\\

\multirow{2}{*}{$Z(10650)$}&1.46  & 1.46 & 1.46  & 1.46  & 1.46  & 1.46  & 1.46  & 1.46  & 1.46  & 1.46  & 1.46  & 1.46 & 1.46 &\multirow{2}{*}{-}\\

& - 2.0i& 0.0i& – 1.7i& + 0.5i & - 1.5i & + 1.3i & - 1.3i & + 2.0i & - 1.1i & + 3.0i & - 0.67i &  + 8.0i &0.0i\\

$Z^{+}(4430)$ & 4.81 & 1.46 & 4.44 & 0.62 & 3.97 & -0.77 & 3.69 & -1.89 & 3.38 & -3.56 & 2.58 & -11.94 & 1.46 & - \\

$D_{s0}^{\pm}(2317)$ & 1.37 & 0.26 & 1.25 & -0.01 & 1.1 & -0.48 & 1 & -0.85 & 0.9 & -1.4 & 0.63 & -4.17 & 0.26 &- \\

$D_{s1}^{\pm}(2460)$ & 2.56 & 1.46 & 2.43 & 1.19 & 2.28 & 0.73 & 2.19 & 0.37 & 2.09 & -0.18 & 1.83 & -2.91 & 1.46 &-\\

\\
\hline
\hline
\end{tabular}
}
\end{table*}
\end{center}

\begin{table*}[]
\begin{center}
\caption{\small The scattering length $a_{s}$ and effective range $r_{e}$ are calculated for di-mesonic systems by using Eq.(\ref{as and re equation}) for different values of renormalization constant Z. The values of $a_{s}$ and $r_{e}$ are in fm. The calculated binding energies are considered which are tabulated in Table-\ref{threshold and expected binding energy}.}
\label{as and re for di-mesonic states}
\scalebox{0.85}{
\begin{tabular}{cccccccccccccccc}
\hline
I$(J^{P})$ & State   &\multicolumn{2}{c}{Z=0} &\multicolumn{2}{c}{Z=0.2} &\multicolumn{2}{c}{Z=0.4} &\multicolumn{2}{c}{Z=0.5} &\multicolumn{2}{c}{Z=0.6} &\multicolumn{2}{c}{Z=0.9} &\multicolumn{2}{c}{Z=1} \\
\cline{3-16}
 & &$a_{s}$ &$r_{e}$ &$a_{s}$ &$r_{e}$& $a_{s}$ &$r_{e}$ &$a_{s}$ &$r_{e}$ & $a_{s}$ &$r_{e}$ & $a_{s}$ &$r_{e}$ & $a_{s}$ &$r_{e}$\\
\hline
\hline
\\
$0(1^{+})$ & $pn$ &5.78 & 1.46 & 5.3 & 0.38 & 4.7 & -1.42 & 4.34 & -2.86 & 3.93 & -5.01 & 2.9 & -15.81 & 1.46 &-\\
$0(0^{+})$&$K\overline{K}$ & 3.83	&	0.26	&	3.43	&	-0.63	&	2.94	&	-2.11	&	2.64	&	-3.3	&	2.3	&	-5.09	&	0.91	&	-31.83	&	0.26 & - \\

$1(0^{+})$&$K\overline{K}$ & 3.36	&	0.36	&	3.03	&	-0.39	&	2.61	&	-1.64	&	2.36	&	-2.65	&	2.08	&	-4.15	&	0.91	&	-26.68	&	0.36	& - \\

$0(0^{+})$&$\rho\overline{\rho}$ & 4.49 & 1.46 & 4.15 & 0.71 & 3.73 & -0.56 & 3.48 & -1.56 & 3.19 & -3.08 & 2.01 & -25.77 & 1.46 & - \\

\multirow{2}{*}{$0(2^{+})$} &\multirow{2}{*}{$\rho\overline{\rho}$} & 1.46 & 1.46 & 1.46 & 1.46 & 1.46 & 1.46 & 1.46 & 1.46 & 1.46 & 1.46 & 1.46 & 1.46 & 1.46 & \multirow{2}{*}{-}\\
&&-112.6i&+0.0i&-100.1i&+28.1i&-84.5i&+75.1i&-75.1i&+112.6i&-64.3i&+169.0i&- 20.4i& +1014.1i& +0.0i \\

$0(0^{+})$&$K^{*}\overline{K^{*}}$ & 3.98 & 1.46 & 3.7 & 0.83 & 3.35 & -0.22 & 3.14 & -1.06 & 2.9 & -2.32 & 1.92 & -21.2 &  1.46 &  -\\

$0(1^{+})$&$D\overline{D^{*}}$ & 4.36 & 1.46 & 4.04 & 0.74 & 3.64 & -0.47 & 3.39 & -1.44 & 3.12 & -2.89 & 1.99 & -24.63 &  1.46 & - \\

$0(2^{+})$&$D^{*}\overline{D^{*}}$ & 3.7 & 1.46 & 3.45 & 0.9 & 3.14 & -0.03 & 2.96 & -0.78 & 2.74 & -1.9 & 1.87 & -18.71 &  1.46 & - \\

$0(1^{+})$&$B\overline{B^{*}}$ & 3.25 & 1.46 & 3.05 & 1.01 & 2.8 & 0.27 & 2.65 & -0.33 & 2.48 & -1.22 & 1.79 & -14.64 &  1.46 & - \\

$0(1^{+})$&$B^{*}\overline{B^{*}}$ & 3.25 & 1.46 & 3.05 & 1.02 & 2.8 & 0.27 & 2.65 & -0.33 & 2.48 & -1.22 & 1.79 & -14.63 &  1.46 & - \\

$0(0^{-})$&$D^{*\pm}\overline{D_{1}^{0}}$& 4.05 & 1.46 & 3.76 & 0.82 & 3.4 & -0.26 & 3.18 & -1.12 & 2.94 & -2.41 & 1.93 & -21.8 &  1.46 & - \\
$0(0^{+})$ & $D^{0}\overline{K^{\pm}}$ & 2.05	&	0.26	&	1.85	&	-0.18	&	1.6	&	-0.93	&	1.45	&	-1.52	&	1.28	&	-2.42	&	0.59	&	-15.81	&	0.26	
& - \\
$0(1^{+})$ & $D^{*0}\overline{K^{\pm}}$ &  2.26	&	0.26	&	2.04	&	-0.24	&	1.76	&	-1.07	&	1.6	&	-1.74	&	1.41	&	-2.74	&	0.63	&	-17.75	&	0.26	&
 -  \\ 
 
\hline
\hline
\end{tabular}
}
\end{center}
\end{table*} 
\section{Results and Discussion}
We have fixed the potential parameters to get experimental binding energy of the deuteron and take all these parameters same for rest of the calculation. The hadron masses, exchange meson coupling constants, regularization parameters are tabulated in Table-\ref{OBEP parameters}. The screen parameter 'c' of the screen Yukawa-like potential is the only free fitting parameter of the model which is also fitted only for deuteron, and fixed for rest of the calculations.

For c=0.0686 GeV, we have obtained the calculated binding energy (2.221 MeV) of the deuteron in agreement with empirical  value (2.224 MeV), see Table-\ref{threshold and expected binding energy}.   

In Fig-\ref{s-wave obe plots}, the graphs of effective s-wave OBE potential are plotted for attempted di-mesonic states. We can see from fig-\ref{s-wave obe plots} that the s-wave OBE potential is repulsive at short range in all cases. The spin-isospin  (S,I) = (0,0), (0,1) and (1,0) channels are attractive. However,  the strength of these attractive s-wave OBE potential with the respective spin-isospin channel are very shallow. Moreover, this attractive strength appeared at very far range (between 5 fm to 10 fm). Therefore, although there is an attractive strength of s-wave OBE potential, it is difficult to get a bound state with pure s-wave OBE interaction. The strength of effective s-wave OBE is vary shallow due to very delicate cancellation of the individual meson exchange with each others.
Hence, the Yukawa-like screen potential is used to get additional strength for net effective potential. As a consequence,
screen Yukawa-like potential shows large impact on the net effective interaction potential, where this potential is sensitive to the parameter 'c'.

The threshold mass, reduced mass and natural energy scale for molecular interpretation of attempted dimesonic states are tabulated in Table-\ref{natural energy scale}. The calculated results of the di-mesonic states by using the effective interaction potential are tabulated in Table-\ref{threshold and expected binding energy}.  The results of the scattering length ($a_{s}$) and effective range ($r_{e}$) are obtained by using Eq(\ref{as and re equation}). By using the expected binding, the results of $a_{s}$ and $r_{e}$ for the dimesonic states depicted in Table-\ref{threshold and expected binding energy} are shown in Table-\ref{expected as and re from exp BE} while by using the calculated binding energy, the results are shown in Table-\ref{as and re for di-mesonic states}.

We have used the positive binding energy ($\epsilon$) for the calculation of $a_{s}$ and $r_{e}$, where the bound state is located at $\epsilon = - \epsilon$. The range correction ${\cal{O}}(1/\beta)$, where $\beta$ is the inverse range of the force,  is approximated as $1/m_{\pi}$, where $m_{\pi}$ is the mass of the $\pi$-meson which is lightest  exchange meson in s-wave OBE.  While for those systems,  in which the pion exchange is not allowed, the ${\cal{O}}(1/\beta)$ is approximated as $1/m_{\sigma}$, where $m_{\sigma}$ is the mass of the $\sigma$-meson.

The presented results in this article are discussed on the basis of obtained binding energy compared to expected binding energy,  Low-energy Universality theorem \cite{Braaten-prd2004-69}, and natural energy scale of di-hadronic systems which could be estimated by $m_{\pi}^{2}/2\mu$, where $\mu$ is the reduced mass of two hadrons \cite{Braaten-prd2004-69}.     

Eric Braaten \cite{Braaten-prd2004-69} explained the Low-energy Universality for X(3872) very near to the threshold and large scattering length. Braaten was defined the Low-energy Universality as: the low energy few-body observables for non-relativistic particles with short-range interactions and a large scattering length have universal features that are insensitive to the details of the mechanism that generates the large scattering length. Thus, if $a_{s}>0$ then it predict a shallow two body bound state. However, a shallow s-wave bound state leads to scattering length large compared to the natural length scale $1/m_{\pi}$ and it implies that the probability for molecular interpretation increases as scattering length increases \cite{Braaten-prd2004-69}. Weinberg \cite{Weinberg-pr1965-137} pointed out that for any elementary state, the value of $a_{s}$ would be less than $R=1/\sqrt{2\mu \epsilon}$ (size of the molecule) while $r_{e}$ would be large and negative. Thus, for any composite system  $r_{e}$ should be small and positive rather than large and negative. 

Deuteron is the only known and strong candidate of hadronic molecule whose experimental scattering length is found large and positive which is well excepted on the both theoretical and experimental foreground, whereas the other possible candidates are still in debate. On the other hand, the binding energy of deuteron is very small (2.22 MeV) to the natural energy scale which is about 20 MeV. With the expected value of binding energy (2.224 MeV) of the deuteron, the scattering length and effective range are in agreement with experimental values ($a_{s}$=5.41 fm, $r_{e}$=1.75 fm) for Z = 0 or 0.1, whereas effective range ($r_{e}$) gains negative value from Z$>$0.3 and become large as $Z\rightarrow 1$. As mentioned by Weinberg, for Z$\rightarrow$0 the state is bringing the composite structure and the experimental values of $a_{s}$ and $r_{e}$ are matched for the Z$\simeq$0 which indicate that the deuteron is in a composite state. 

For state $f_{0}(980)$, the calculated binding energy is in agreement with the expected binding energy, whereas for the state $a_{0}(980)$ calculated binding energy is underestimated. In both cases only the $\sigma$, $a_{0}$, and $\omega$ exchange contributes to net s-wave OBE potential and thus only these meson exchange are considered. The deep binding energy is leading to the large scattering length. The natural energy scale  for the states $f_{0}(980)$ and $a_{0}(980)$ is about 36 MeV if they have $K\overline{K}$ in their internal structure. Both the expected as well as calculated binding energy are very small from natural energy scale which leading the probability of the deuteron-like structure. 

The state $f^{'}_{2}(1525)$ is found unbound or almost on threshold with +0.003 MeV binding energy in the present calculation. If $f^{'}_{2}(1525)$ has $\rho \overline{\rho}$ in its internal structure then the natural energy scale gets about 25 MeV, whereas the expected binding energy is very close to its natural energy scale, thus, with this deep binding $f^{'}_{2}(1525)$ gain large positive scattering length and effective range from Z=0 to Z=0.5. 

For the states $f_{0}(1500)$ and $f_{0}(1710)$, the natural energy scale are about 25 MeV and 20 MeV, if they have $\rho \overline{\rho}$ and $K^{*}\overline{K^{*}}$ in their substructure, respectively. However, the expected binding energy for both states are well above the natural energy scale, see Table-\ref{threshold and expected binding energy}. Definitely, the deep binding lead positive effective range, but these binding energies are above the natural energy scale of the molecular interpretation and thus it is indicating the some other substructure for these states.

The state X(3872) is good example of very near threshold structure. The state X(3872) have been extensively studied as $D\overline{D^{*}}$($\overline{D}D^{*}$) molecule and it is lying just below the $D\overline{D^{*}}$ threshold. The binding energy of the X(3872) (0.16 MeV) is very small from its natural energy scale which is about 10 MeV if it consist  $D\overline{D^{*}}$ in its internal structure. With the expected binding energy (0.16 MeV) of the state, one get the large positive value of $a_{s}$ but negative $r_e$ from Z$>$0.2, see Table-\ref{expected as and re from exp BE}. Hence, if X(3872) is have a dominated or a pure molecular structure then the value of Z must be less than 0.2, while with the calculated binding energy which is overestimated from experimental value, $r_e$ becomes negative for Z$>$0.3.

The state $Z(4430)^{+}$ for which the expected binding energy is (1.58 MeV) tends to $a_{s}$=4.8 fm and $r_{e}$ =1.4 fm for Z=0 (pure molecule) while $r_{e}$ becomes negative and large for Z$>$0.3, whereas with the calculated binding energy $r_e$ is getting negative from Z$>$0.3. Guo-Zhan Meng in \cite{Meng-prd2009-80} has studied low energy scattering of $D^{*}$ and $D_{1}$ meson using quenched lattice QCD and reported scattering length $a_{s}$=2.52 fm and effective range $r_{e}$=0.7 fm in $J^{P}=0^{-}$ channel. The small and positive effective range is lead to molecular interpretation of the state $Z(4430)^{+}$ composed of $D^{*}\overline{D_{1}}$.

For the state $D_{s0}(2317)^{\pm}$ and $D_{s1}(2460)^{\pm}$ calculated binding energies are underestimated compared with expected binding energies. In the case of $D_{s0}(2317)^{\pm}$, the $\sigma$, $a_{0}$, and $\omega$ exchange contributes to net s-wave OBE potential, thus, only these meson exchange are considered. The range correction ${\cal{O}}(1/\beta)$ is considered as $m_{\sigma}^{-1}$. The natural energy scale for both states are near 25 MeV and the expected binding energies of both states are well above the natural energy scale. However, the deep expected binding energies leads positive scattering length. In the case of  $D_{s0}(2317)^{\pm}$ the effective range get negative value from Z=0.2 while in the case of $D_{s1}(2460)^{\pm}$ effective range get negative value for Z$>$0.5.  

In Ref. \cite{Mohler-prl2013-111,Mohler-prd2013-87}, authors studied the lattice QCD simulation  for the$D_{s0}(2317)^{\pm}$ near to DK threshold and reported the scattering length and effective range $1.33\pm 0.20$ fm and $0.27\pm 0.17$ fm, respectively, at pion mass, $m_{\pi}=156 MeV$. Whereas, the Weinberg compositeness theorem predicts $a_{s}=1.37$ fm and $r_{e}=0.26$ fm for purely molecular state (Z=0) (where the inverse $\rho$ mass was assumed for the range of the forces and same is assumed in present work). The same comparable  results were reported in ref. \cite{L.Liu-prd2013-87} by using unitarrized chiral perturbation theory. Both of the studies \cite{Mohler-prl2013-111,L.Liu-prd2013-87} were predicted $D_{s0}(2317)^{\pm}$ as a DK molecule.    

\section{Summary and Conclusion} 
In summary, we have attempted two interesting 
challenges of the hadronic molecular model (i) inter hadronic interaction within hadronic molecule (ii) identification of the hadronic molecules.

It is well known that hadronic molecules should be near to s-wave threshold. Therefor, we have incorporated the s-wave one boson exchange interaction potential. we have observed that the attractive strength of s-wave OBE potential is very shallow. Therefor, the screen Yukawa-like potential used along with OBE potential to get additional attractive strength. We have successfully generated the mass spectra of di-mesonic states where the model parameters have been fixed for deuteron.   

We have discussed the compositeness theorem  which provides the observables for identification of the hadronic molecules in terms of scattering length and effective range.

In order to determine the nature of the state, the values of scattering length ($a_{s}$) and effective range ($r_{e}$) obtained by using compositeness theorem (Eq.(\ref{as and re equation})) and presented in Table-\ref{expected as and re from exp BE}, \ref{as and re for di-mesonic states} are  became to be useful only if one have data of $a_{s}$ and $r_{e}$ by some other means for comparison, i.e. either from experiment or from lattice QCD. Such effort can be attempted in lattice QCD by using the L$\ddot{u}$scher formalism \cite{luscher-cmp1986-105,Luscher-npb1991-531}. In L$\ddot{u}$scher formalism, measuring the low-energy scattering observables in lattice QCD, we can extract the comositeness by using Eq(\ref{as and re equation}) ( also see for review \cite{Guo-RevModPhys2018-90}). Some efforts have been made in Refs.\cite{Martinez-prd2012-85,Ozaki-prd2013-87,Albaladejo-prd2013-88,Agadjanov-jhep2015-2015}. In Ref. \cite{Agadjanov-jhep2015-2015}, authors proposed that the scattering amplitude in the finite volume can be obtained from the corresponding loop function and  mentioned that the use of partially twisted boundary conditions is easier than studying the volume dependence in lattice for measuring the compositeness. Indeed, it is necessary to get access to the scattering length and the effective range by some other tools like lattice QCD data or experimental data  to get some imprints about the nature of the state by using the results presented in this study.

From these analysis, with the calculated results of the present formalism, the states pn bound state, $f_{0}(980)$, $a_{0}(980)$, X(3872), $Z(4430)^{+}$ are appeared as the strong hadronic molecular candidates. Exceptionally, the state X(3872) is needed very fine tuning of parameters to get empirical binding energy. Therefor, the state X(3872) has seemed to driven dominant molecular structure rather a pure molecule, similar results was found in our previous study \cite{Rathaud-epjp2017-132}. Moreover, we have predicted spin-2 charm partner of X(3872) as a $X_{2c}(4013)$. The charged bottomonium like states Z(10610) and Z(10650) have not identified as $B\overline{B^{*}}$ and $B^{*}\overline{B^{*}}$ hadronic molecules, respectively as they are just above the threshold. But we have found bound states of $B\overline{B^{*}}$ and $B^{*}\overline{B^{*}}$. In contrast to these Z states, the states $f_{0}(1500)$ and $f_{0}(1710)$ have needed very deep binding which has fall out of the natural energy scale of the shallow bound state interpretation. However, decay results for the $f_{0}(1710)$ in our previous study \cite{Rai-epjc2015} has favored the molecular structure to $f_{0}(1710)$. But in overall scenario, we have not reach on any conclusion regarding the substructure of $f_{0}(1500)$, $f_{0}(1710)$.

In conclusion, we strongly suggest that there must be dipole like interaction between two color neutral states, as in the present study we have found the dominance of the Yukawa screen-like potential over the s-wave OBE. This type of interaction could explore the certain decays of the particles in which the stability of the molecule will be given by the shorter lived component.  To determine whether all these states have a molecular or a confined structure or something else, we need a reasonable estimates of scattering length and effective range  from experimental and lattice QCD data, to reach on a strong conclusion for the identification of substructure of all these states.

 
The aim of the present article is to explore the interaction between two hadrons and their identification as a molecule. With our proposed interaction, we have found di-mesonic bound state in attractive spin-isospin channels with dominating screen type interaction. With same model, the results of mass spectra of meson-baryon and di-baryonic molecules  along with detail analysis of meson exchange potential will present in succeeding   
publication. 


\end{document}